\documentclass[twocolumn,superscriptaddress,nofootinbib]{revtex4-1}

\usepackage[utf8]{inputenc} 
\usepackage[T1]{fontenc}    
\usepackage{hyperref}       
\usepackage{url}            
\usepackage{booktabs}       
\usepackage{amsfonts}       
\usepackage{nicefrac}       
\usepackage{microtype}      
\usepackage{lipsum}
\usepackage{graphicx}
\usepackage{amsmath}
\usepackage{float}

\usepackage[table]{xcolor}
\usepackage{color}
\usepackage{colortbl}
\usepackage{multirow}

\definecolor{midgreen}{rgb}{0.52, 0.73, 0.4}

\graphicspath{ {figures/} }

\begin{document}

\title{Building a testable shear viscosity across the QCD phase diagram}

\author{Emma McLaughlin}
\affiliation{Department of Physics, Columbia University, 538 West 120th Street, New York, NY 10027, USA}
\author{Jacob Rose}
\affiliation{Department of Physics,
Dr. Karl-Remeis Observatory  - Astronomical Institute,
Bamberg D-96049, Germany}
\author{Travis Dore}
\affiliation{Illinois Center for Advanced Studies of the Universe, Department of Physics, University of Illinois at Urbana-Champaign, Urbana, IL 61801, USA}
\author{Paolo Parotto}
\affiliation{University of Wuppertal, Department of Physics, Wuppertal D-42097, Germany}
\author{Claudia Ratti}
\affiliation{Department of Physics, University of Houston, Houston, TX  77204, USA}
\author{Jacquelyn Noronha-Hostler}
\affiliation{Illinois Center for Advanced Studies of the Universe, Department of Physics, University of Illinois at Urbana-Champaign, Urbana, IL 61801, USA}

\begin{abstract}
Current experiments at the Relativistic Heavy Ion Collider (RHIC) are probing finite baryon densities where the shear viscosity to enthalpy ratio $\eta T/w$ of the Quark Gluon Plasma remains unknown. We use the Hadron Resonance Gas (HRG) model with the most up-to-date hadron list to calculate $\eta T/w$ at low temperatures and at finite baryon densities $\rho_B$. We then match $\eta T/w$ to a QCD-based shear viscosity calculation within the deconfined phase to create a table across $\left\{T,\mu_B\right\}$ for different cross-over and critical point scenarios  at a specified location. We find that these new $\eta T/w(T,\mu_B)$ values would require initial conditions at significantly larger $\rho_B$, compared to ideal hydrodynamic trajectories, in order to reach the same freeze-out point.
\end{abstract}
 

\maketitle

\section{Introduction}
Relativistic heavy-ion collisions at the Large Hadron Collider (LHC) and Relativistic Heavy-Ion Collider (RHIC) have successfully reproduced the phase transition from the Quark Gluon Plasma into a hadron resonance gas in the laboratory.  Since 2006 it has been understood that this phase transition is a smooth cross-over \cite{Aoki:2006br,Borsanyi:2010cj,Borsanyi:2010bp,Bazavov:2014pvz,Borsanyi:2020fev,Ratti:2018ksb,Guenther:2020vqg}. For this reason, the definition of the pseudo-critical temperature $T_{pc}$ is not unique, but rather depends on the observable
considered. If Quantum Chromodynamics (QCD) exhibits a critical point at large baryon densities, one expects the different pseudo-critical temperatures to converge to the critical temperature $T_c$ \cite{Endrodi:2011gv,Rougemont:2017tlu,1848212,Bellwied:2018tkc,Noronha-Hostler:2016rpd} at the critical point. However, it also possible that there is no critical point 
and only a cross-over is seen at larger baryon densities. 

One crucial achievement of heavy-ion collisions has been the discovery that the Quark Gluon Plasma is a nearly perfect fluid with a shear viscosity to entropy density ratio $\eta/s$ which at $\mu_B=0$ is
close to the KSS limit \cite{Kovtun:2004de} (that we now know is no longer a strict bound \cite{Kats:2007mq,Brigante:2007nu,Brigante:2008gz,Buchel:2008vz,Brigante:2008gz,Critelli:2014kra,Finazzo:2016mhm}). Due to the extremely short lifetime of the produced Quark Gluon Plasma, relativistic viscous hydrodynamic calculations have been vital to understand the dynamics of this cross-over phase transition. It is common practice in the field to use $\eta/s$ as a free parameter, and extract its value from experimental data by comparing flow harmonics to theoretical predictions from relativistic hydrodynamics \cite{Romatschke:2007mq,Bozek:2011ua,Heinz:2013th,Luzum:2013yya,Niemi:2015qia,Noronha-Hostler:2015uye,McDonald:2016vlt,Bernhard:2019bmu,Alba:2017hhe}. However, since there are many other free parameters that influence the extraction of $\eta/s$, this is not a trivial task.

Ideally, $\eta/s$ would be calculated directly from lattice QCD but this turns out to be an ill-posed problem, as it involves inversion methods over a discrete set of lattice points for the correlator of the energy-momentum tensor. Besides, the signal for this specific correlator is dominated by the high-$\omega$ part of the spectral function 
\cite{Aarts:2002cc,Pasztor:2018yae}, 
which makes the inversion even harder. For these reasons, one must turn to alternative approaches such as the hadron resonance gas (HRG) model \cite{NoronhaHostler:2008ju,Pal:2010es,Khvorostukhin:2010aj,Tawfik:2010mb,Alba:2015iva,Ratti:2010kj,Tiwari:2011km,NoronhaHostler:2012ug,Kadam:2014cua,Kadam:2015xsa,Kadam:2015fza,Kadam:2018hdo,Mohapatra:2019mcl}, transport theory \cite{Wesp:2011yy,Ozvenchuk:2012kh,Rose:2017bjz,Rais:2019chb}, holography \cite{Kovtun:2004de}, Color String Percolation Model \cite{Sahu:2020mzo}, linear sigma model \cite{Heffernan:2020zcf} or QCD-motivated alternatives \cite{Haas:2013hpa,Christiansen:2014ypa,Dubla:2018czx,Ghiglieri:2018dib}, to name a few.  These calculations are often only performed at $\mu_B=0$. Hence, they would be appropriate primarily for LHC and top RHIC energies, but not for large baryon densities.

While calculations currently exist from some of these models at finite baryon densities, there are difficulties in systematically studying $\eta T/w(T,\mu_B)$\footnote{Note that at finite $\mu_B$ the enthalpy $w=\varepsilon+p$ is used rather than entropy} for different types of phase transitions (cross-over vs. a critical point within the dynamical H or B university classes) since some of these models are only applicable within the hadronic phase (e.g. HRG model \cite{Denicol:2013nua,Kadam:2014cua}, UrQMD \cite{Demir:2008tr}, and SMASH \cite{Rose:2017bjz} calculations), or they are beholden to dynamic university class of the model itself (e.g. holography in Universality class B \cite{Rougemont:2017tlu,Critelli:2017oub}, which is also limited to a constant $\eta/s(T,\mu_B)=0.08$ for current studies at finite $\mu_B$ \footnote{Alternatives to $\eta/s(T,\mu_B=0)=0.08$ exist \cite{Kats:2007mq,Brigante:2007nu,Brigante:2008gz,Buchel:2008vz} at vanishing baryon densities where an action is formulated that allows for derivatives up to the 4th order. However, these calculations have not yet been incorporated into non-conformal AdS. Since non-conformality is vital for understanding the QCD phase transition, the current framework cannot provide non-trivial information about $\eta/s(T,\mu_B)$ at a critical point.}). Furthermore, at larger baryon densities repulsive interactions have been shown to be more relevant \cite{Vovchenko:2016rkn,Huovinen:2017ogf}, which should be taken into account in the hadronic phase. 

In this paper we establish a framework for combining $\eta T/w(T,\mu_B)$ in the confined phase -- from an interacting hadron resonance gas (HRG) -- and in the deconfined phase -- with a QCD-based assumption -- in the presence of a cross-over phase transition.  The key element is that we use the finite $\mu_B$ behavior from the HRG model to construct $\eta T/w(T,\mu_B)$ across the phase diagram relevant to heavy-ion collisions.  We note that the framework is generic enough to be easily updated if more realistic QCD-based calculations become available at finite $\mu_B$. We use an interacting hadron resonance gas with the most up-to-date particle resonance list from the Particle Data Group (PDG16+) \cite{Alba:2017mqu}, which was shown to be a reasonable fit compared to lattice QCD data, can describe net-Kaon fluctuations \cite{Bellwied:2018tkc}, off-diagonal susceptibilities \cite{Bellwied:2019pxh}, thermal fits  \cite{Alba:2020jir},  and  works well within relativistic viscous hydrodynamic calculations \cite{Alba:2017hhe}. We base the treatment of the  deconfined phase on the parametrized version of $\eta/s(T,\mu_B=0)$ from Ref.~\cite{Dubla:2018czx}, which is adjusted to match the hadron resonance gas model at the phase transition assuming a minimum value $\eta/s\sim 0.08$. In this work we construct four different profiles for $\eta T/w(T,\mu_B)$: we first consider the cases of a smooth or sharp crossover, then the case with a critical point, first at $\left\{T, \mu_B\right\} = \left\{143, 350\right\}$ MeV -- to match to the publicly available BEST collaboration EoS \cite{Parotto:2018pwx}, then at $\left\{T, \mu_B\right\} = \left\{89, 724\right\}$ MeV -- matching the prediction from holography in Ref.~\cite{Critelli:2017oub}\footnote{We remind the reader that, while the result in Ref. \cite{Critelli:2017oub} is a prediction on the location of the critical point, the BEST collaboration EoS allows the user to pick its location, and the one mentioned here is the one used in Ref. \cite{Parotto:2018pwx} for illustration purposes.}.


\section{Hadron Resonance Gas Model}

With the HRG model one can calculate the pressure, energy density and total particle density of species $i$ assuming that the hadrons are point like particles:
\begin{widetext}
\begin{eqnarray}
    \frac{p(T,\mu_B,\mu_S,\mu_Q)}{T^{4}} &=&  \sum\limits_{i}(-1)^{B_{i}+1} \frac{g_{i}}{2\pi^{2}}\int\limits_{0}^{\infty} p^{2}\ln{[1+(-1)^{B_{i}+1}e^{(-\frac{\sqrt{p^{2}+m_i^{2}}}{T}+ \tilde{\mu}_i)}]} dp\label{eqn:hrgp}\\
   \frac{ \epsilon (T,\mu_B,\mu_S,\mu_Q)}{T^4}&=&  \sum\limits_{i}\frac{g_{i}}{2\pi^{2}}\int\limits_{0}^{\infty} \frac{p^{2}\sqrt{p^{2}+m_i^{2}}}{(-1)^{B_{i}+1} + e^{(-\frac{\sqrt{p^{2}+m_i^{2}}}{T}+ \tilde{\mu}_i)}} dp\label{eqn:hrge}\\
   \frac{ n_i (T,\mu_B,\mu_S,\mu_Q)}{T^3}&=&  \frac{g_i}{2\pi^2}\int\limits_{0}^{\infty}p^2 \left[ \exp \left(\frac{\sqrt{p^2+m_i^2}}{T}-\tilde{\mu}_i\right)+ (-1)^{B_i-1}\right]^{-1} dp\label{eqn:hrgn}
\end{eqnarray}
\end{widetext}
where
\begin{equation}
    \tilde{\mu}_i\equiv B_i\mu_{B}/T+S_i\mu_{S}/T+Q_i\mu_{Q}/T
\end{equation}
and $g_i$ is the degeneracy of each hadron, $m_i$ is the mass, and $B_i$, $S_i$, and $Q_i$ are the baryon number, strangeness and electric charge carried by each hadron.  Additionally, $\mu_B$, $\mu_S$, and $\mu_Q$ are the corresponding chemical potentials for each conserved charge. 

The other thermodynamic quantities follow 
\begin{eqnarray}
 s(T,\mu_B,\mu_S,\mu_Q)&=& \frac{\partial p (T,\mu_B,\mu_S,\mu_Q)}{\partial T} \nonumber\\
\frac{ \rho_i(T,\mu_B,\mu_S,\mu_Q)}{T^3}&=& \frac{\partial p(T,\mu_B,\mu_S,\mu_Q)/T^4}{\partial (\mu_i/T)}\nonumber\\
 p+\epsilon& =& sT+\sum_{i=B,S,Q}\mu_{i}\rho_{i}\label{eqn:therm}
\end{eqnarray}
where $s$ is the entropy, 
$\rho_i$ where $i=B,S,Q $ is the net density, and $\epsilon$ is the energy density. Note that for the rest of the paper we will abbreviate $\mu_B,\mu_S,\mu_Q\Rightarrow\mu_i$. 

In this paper we first compare two different lists of hadrons from the Particle Data Group, one from 2005 (PDG05) and another developed in Ref.~\cite{Alba:2017mqu} from 2016 that includes all *-**** states (PDG16+).

\subsection{Excluded Volume}

One method for taking into account repulsive interactions is the excluded volume approach \cite{Rischke:1991ke}  wherein each hadron is delegated a volume $v$, and the excluded volume pressure $p_v$ can then be written as
\begin{equation}
 \frac{p_v(T,\mu_i)}{T}= n(T,\mu_i) \exp(-vp_v(T,\mu_i)/T),
\label{vdwpressure}
\end{equation}
which can be solved analytically using the Lambert W function
\begin{equation}
 p_v(T,\mu_i)= \frac{T}{v}W(v\, n(T,\mu_i))\,,
\label{vdwpressuresolution2}
\end{equation}
where $n(T,\mu_i)=\sum_i n_i(T,\mu_i)$ is defined in Eq.\ (\ref{eqn:hrgn}). The remaining thermodynamic quantities can be obtained using the thermodynamic relationships from Eq.\ (\ref{eqn:therm}) and are denoted with a subscript $v$ for excluded volume.
Here we are assuming that all particles have the same volume.  In fact, one could relax that assumption as in Ref.~\cite{Albright:2014gva} or even consider a multicomponent van der Waals \cite{Vovchenko:2017zpj}, but we leave this for future work.

 We obtain the effective hard-core volume from $v= 4\cdot4\pi r^3/3$, where $r$ is the effective core radius. The point-like thermodynamic properties in Eqs.\ (\ref{eqn:hrgp}-\ref{eqn:therm}) can be then restored with $r\rightarrow 0$. 
\begin{figure}
\begin{center}
\includegraphics[width=\linewidth]{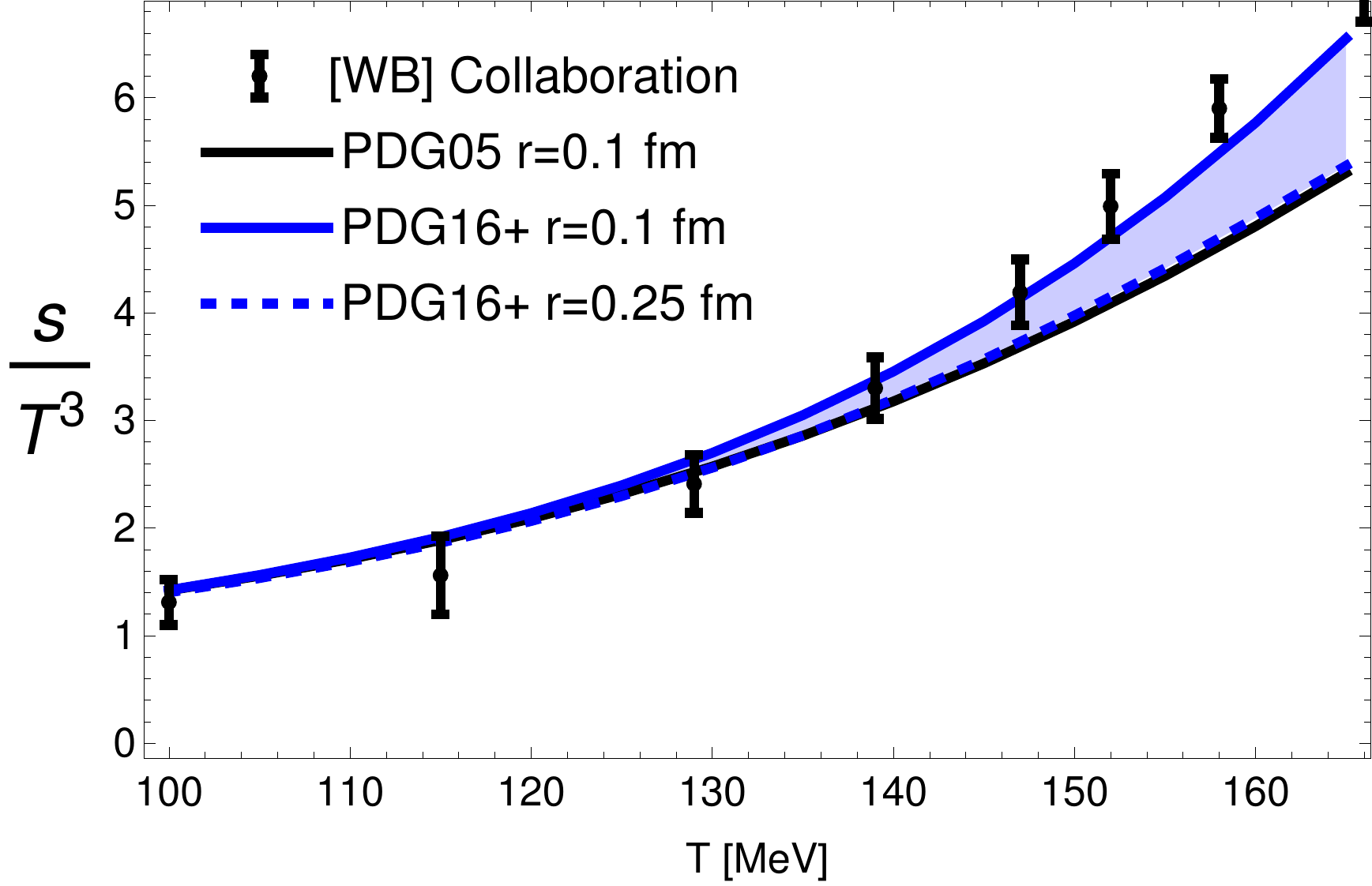}
\caption{
Comparison of the PDDG05 and PDG16+ list with various excluded volumes to $s/T^3$ results from lattice QCD \cite{Borsanyi:2010cj}. The band for the PDG16+ denotes the minimum and maximum $r$ that fits within constraints from Lattice QCD.
} 
\label{fig:shardcore}
\end{center}
\end{figure}

\begin{figure}
    \centering
    \includegraphics[width=\linewidth]{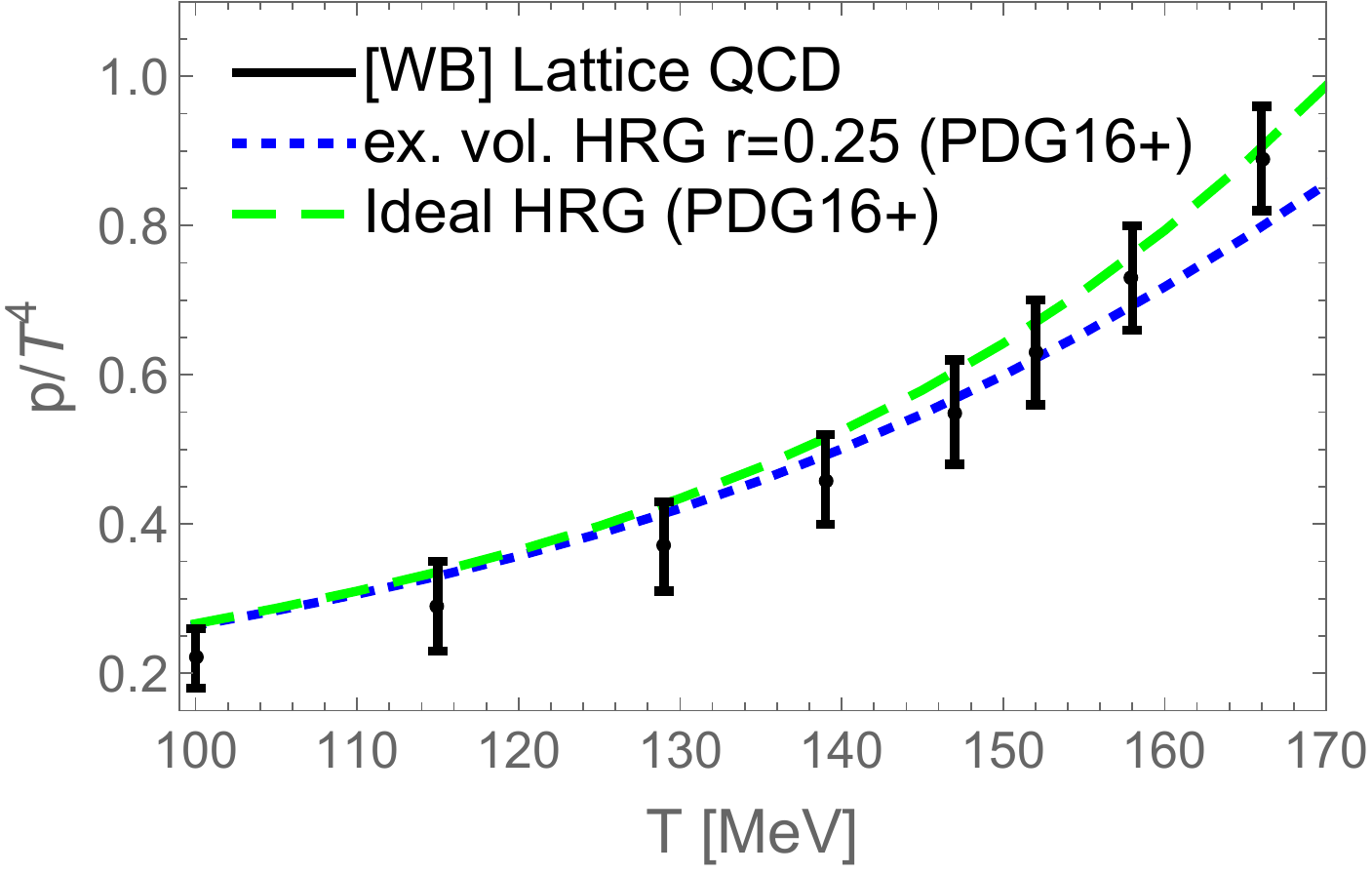}
    \caption{ Pressure comparison to
    lattice QCD results from \cite{Borsanyi:2010cj} with an ideal vs. excluded volume HRG model. 
    }\label{fig:pressure}
\end{figure}

In the excluded volume calculations, there is always the issue of constraining the volume size, since it is an unconstrained free parameter.  In an attempt to constrain the hard core radius, we vary $r$ such that is fits within lattice QCD error-bars up to $T\sim 150$ MeV using the entropy from lattice QCD results shown in Fig.\ \ref{fig:shardcore} for the entropy and in Fig.\ \ref{fig:pressure} for the pressure.  We find that, for the PDG05, there are so few resonances that the largest possible excluded volume that we can use corresponds to a radius $r=0.1$ fm, whereas for the PDG16+ we have significantly more resonances and we can accommodate an excluded volume with a radius up to $r=0.25$ fm.  Beyond this value, we can no longer reproduce the lattice QCD results within error bars. A quick comment on this result compared to previous papers \cite{Gorenstein:2007mw,NoronhaHostler:2012ug} is that here we use the most up-to-date lattice QCD results, whereas in previous work \cite{Gorenstein:2007mw} the lattice QCD results were not yet in the continuum limit and were, therefore, much lower in the confined phase (see e.g., Fig 18 from \cite{Borsanyi:2010cj}) yielding a larger extracted excluded volume.  With current continuum extrapolated lattice results and the most up-to-date particle list, we do not find a need for a large excluded volume. Additionally, in Ref.~\cite{NoronhaHostler:2012ug} Hagedorn resonances were added beyond the PDG, which required a larger excluded volume. This result still holds: if more resonances were eventually measured, a larger excluded would become necessary. While in this work we only explore a single excluded volume, other types of interactions and/or volumes that vary with the hadron mass or flavor content \cite{Alba:2016hwx} would lead to different conclusions and may be relevant at large $\mu_B$. We leave this extension for future work.
\subsection{Shear Viscosity}

\begin{figure}
\begin{center}\includegraphics[width=\linewidth]{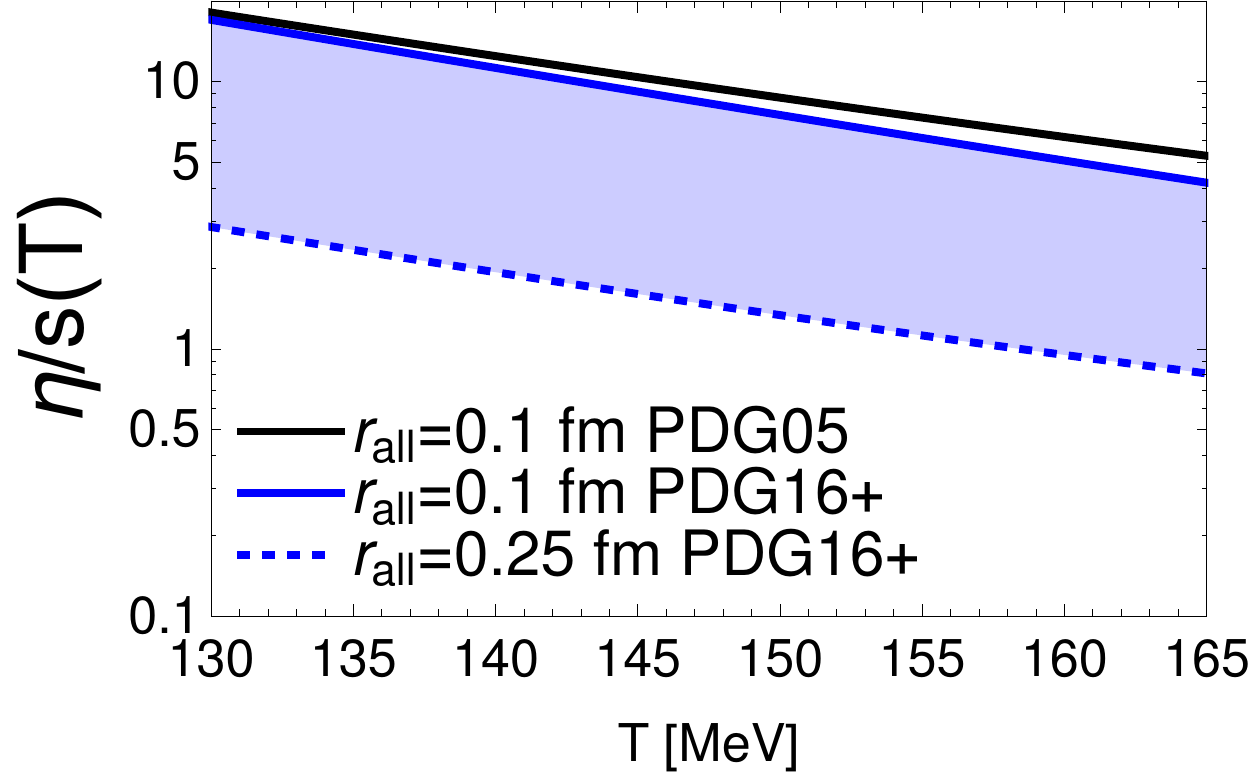}
\caption{
Ratio $\eta/s$ as a function of the temperature. The excluded volume result for the PDG05 list and $r=0.1$ fm (black line) is compared to the results obtained with the PDG16+ list with $r=0.1$ fm (blue solid line) and $r=0.25$ fm (blue dashed line).
} 
\label{fig:etsALL}
\end{center}
\end{figure}

At vanishing baryon density, an excluded volume description has been used in Refs.~\cite{Gorenstein:2007mw,NoronhaHostler:2012ug} to estimate $\eta(T)$ within the hadronic phase.  We extend here this formalism to finite $\mu_B$:
\begin{equation}\label{eqn:etqHRG}
    \eta^{HRG} = \frac{5}{64\sqrt{8}}\frac{1}{r^{2}}\frac{1}{n}\sum\limits_{i}n_{i}\frac{\int\limits_{0}^{\infty}p^{3}e^{-\sqrt{p^{2}+m_{i}^{2}}/T+\tilde{\mu}_i}dp}{\int\limits_{0}^{\infty}p^{2}e^{-\sqrt{p^{2}+m_{i}^{2}}/T+\tilde{\mu}_i}dp} \, \, ,
\end{equation}
assuming -- as mentioned earlier -- a single radius for all species. Note that, in this case, the following relationship between number densities holds:
\begin{equation}
    \frac{n_{i,v}}{n_v}=\frac{n_{i}}{n} \, \, ,
\end{equation}
i.e., one can exchange the excluded volume number density ratio and the ideal gas number density ratio.  If this assumption is relaxed, $n_{i,v}$ and $n_{v}$ must be used in Eq.\ (\ref{eqn:etqHRG}). 

After constraining the radius $r$, it is possible to calculate $\eta/s$ as shown in Fig.\ \ref{fig:etsALL}. Using the PDG05 list, we find that $\eta/s(T)$ is extremely large and is not remotely able to reach the $1/4\pi$ estimate from the KSS bound \cite{Kovtun:2004de}.  
Because of the larger number of states present in the  PDG16+ list, a significant decrease in $\eta/s(T)$ occurs, which brings it down to $\eta/s(T)\sim 0.8$ at T=165 MeV and $\eta/s(T)\sim 0.34$ at T=199 MeV.  

Our next step is to connect to the deconfined phase $\eta^{QGP}/s$ from Refs.~\cite{Christiansen:2014ypa,Dubla:2018czx}. In order to avoid confusion, we first define the following temperatures:
\begin{itemize}
    \item $T_{min}(\mu_B)$: Pseudo-critical temperature, i.e., the temperature where the minimum of $\eta T/w$ is reached for a fixed $\mu_B$;
    \item $T_{sw}$: Switching temperature where the connection between $\eta^{HRG}/s$ and  $\eta^{QGP}/s$ occurs. This might correspond to a lower temperature $T_{sw}\leq T_{min}$;
    \item $T_{ch} (\mu_B)$: Chiral transition line temperature, which varies with $\mu_B$. At vanishing chemical potential one has $T_{ch,0}=T_{ch} (\mu_B=0)$;
    \item $T_{YM}$: a parameter in the definition of $\eta^{QGP}/s$ from Refs.~\cite{Christiansen:2014ypa,Dubla:2018czx};
    \item $T_c$: critical temperature, in the case where there is a critical point (the critical point has a corresponding $\mu_{B,c}$ as well).
\end{itemize}

Since the value $\eta/s(T)\sim 0.8$ is still larger than what one would expect from comparisons of viscous hydrodynamics to experimental data \cite{Bernhard:2016tnd}, we propose the following solution:
\begin{itemize}
    \item Assume a higher pseudo-critical temperature $T_{min}$ for shear viscosity. This is not an unreasonable assumption in light of Refs.~\cite{Christiansen:2014ypa,Rougemont:2017tlu,Dubla:2018czx,Everett:2020xug}, wherein  the minimum $\eta/s(T)$ is found around $T_{min}\sim 200$ MeV. This leads to a lower $\eta/s(T)$ in the hadronic phase of $\sim0.34$;
    \item Renormalize the minimum to $\eta /s(T_{min})=0.08$;
    \item Utilize the parametrized $\eta/s(T)$ from Ref.~\cite{Dubla:2018czx} above $T_{sw}$, which we also normalize to $\eta/s(T_{min}) = 0.08$.
    Then, in the deconfined phase we have: 
    \begin{equation}\label{eqn:etah}
    \quad  \eta^{QGP}/s =f\left[     \frac{a}{\alpha^{\gamma}_{s,HQ}(cT/T_{YM})}+\frac{b}{(T/T_{YM})^{\delta}}\right] \, \, ,
    \end{equation} \normalsize
    where the function $\alpha_{s,HQ}(z)$ reads:
    \begin{equation}
    \alpha_{s,HQ}(z) = \frac{1}{\beta_{0}}\frac{z^{2}-1}{z^{2}log z^{2}} \, \, ,
    \end{equation}
    the constants $a = 0.2$, $b = 0.15$, $c = 0.79$, and the exponents $\delta = 5.1$, 
    $\gamma = 1.6$, $\beta_0=11-2n_f/3$ (where $n_f$ is the number of flavors), and $f$ is an
    overall normalization constant that we need to ensure the correct normalization at $T=T_{min}$ and $\mu_B=0$.
    \end{itemize}
    This allows us to build a $\eta/s(T)$ at $\mu_B=0$ that has a temperature dependence at least motivated by the HRG model (even if the overall magnitude is not).  

We must also implement a finite $\mu_B$ dependence, which is somewhat non-trivial because it depends on the existence and location of a critical point.  Subtle differences appear when only a cross-over or a critical point are present in the phase diagram.  We also must state quite clearly, that we do \emph{not} incorporate any critical scaling of the shear viscosity within this framework.  However, since the shear viscosity scales as $\xi^{(4-d)/19}$ where $d$ is the number of spatial dimensions (in comparison the bulk viscosity scales as $\xi^3$ \cite{Monnai:2016kud}), we argue that criticality would not considerably affect the shear viscosity.  What in turn will affect the shear viscosity is the exact profile of the line of pseudo-critical temperatures $T_{min}(\mu_B)$, i.e. the location of the minimum of $\eta T/w$ across the phase diagram. For this reason, in the following we consider very different choices depending on whether only a cross-over or a critical point are present.

\subsection{Cross-over}

The implementation of a cross-over phase transition is straightforward. We test two different approaches: a sharp cross-over vs. a smooth cross-over. In both cases we connect to $\eta^{QGP}/s$.

\begin{figure*}
\begin{center}\includegraphics[width=0.8\textwidth,trim=0 20cm 0 0, clip]{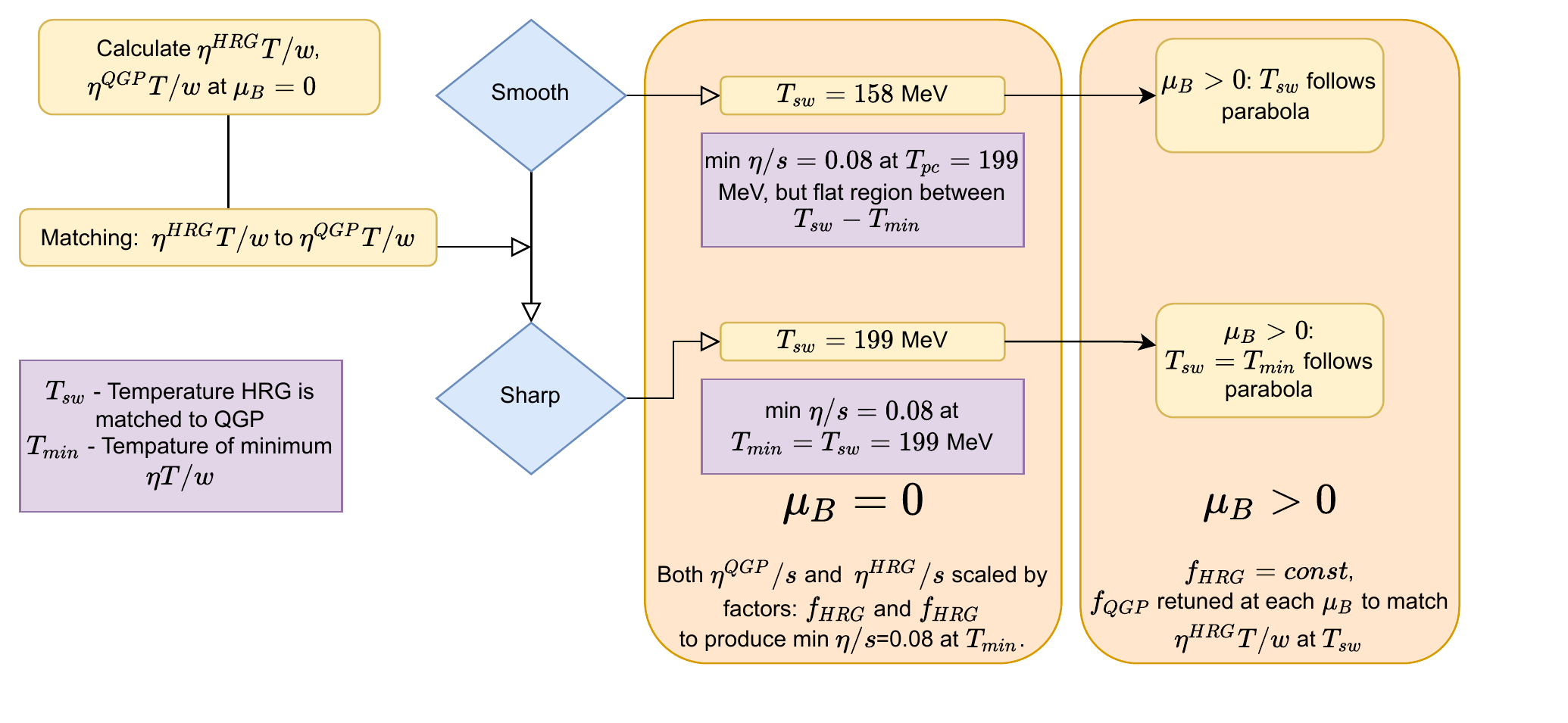}
\caption{
Flow chart of our  $\eta T/w\left\{T,\mu_B\right\}$ procedure for a cross-over. 
} 
\label{fig:flow}
\end{center}
\end{figure*}

One of the main purposes of this work is to construct a temperature-dependent shear viscosity over enthalpy ratio $\eta T/w$ (recall $\eta T/w=\eta T/(\varepsilon+p)$, which simplifies to $\eta/s$  at $\mu_B=0$).  In fact, we would like to construct a shear viscosity that can vary across the entire 4-dimensional QCD phase diagram relevant to heavy-ion collisions, i.e. spanning over $\left\{T,\mu_B,\mu_S,\mu_Q\right\}$.  Unfortunately, such a model does not exist in the QGP phase beyond Ref.~\cite{Rougemont:2017tlu}, although the authors of Ref.~\cite{Christiansen:2014ypa} may eventually incorporate a dependence on the baryon chemical potential, which would be quite useful. For this reason, in this paper we let the behavior of the HRG model drive the $\mu_B$-dependence of the shear viscosity as follows (a flow chart of this procedure is shown in Fig.\ \ref{fig:flow}). We reserve for future work the extension to the full $\left\{\mu_B,\mu_S,\mu_Q\right\}$, which we can anticipate will further influence hydrodynamic trajectories \cite{Monnai:2021kgu}.  

First, we calculate the excluded volume shear viscosity across $\left\{T,\mu_B\right\}$.
The enthalpy is then calculated once again in the HRG model with the mass-dependent volume corrections. With the simple assumption $\mu_S=\mu_Q=0$ one finds that, at large $\mu_B$, there is a suppression in $\eta T/w$, which has already been demonstrated in a number of previous papers \cite{Denicol:2013nua,Kadam:2014cua}. 

As previously discussed, we do not have a  good theory to describe the shear viscosity of the QGP phase across $\left\{T,\mu_B\right\}$. Thus, we simply take the same parameterization form as in Eq.\ (\ref{eqn:etah}), and vary the overall magnitude as well as the switching temperature with $\mu_B$. Everything remains the same as in Eq.\ (\ref{eqn:etah}), except for $T_{YM}$ and $f$. In order to determine $T_{YM}(\mu_B)$ and $f(\mu_B)$ we need to determine the temperature at which the matching of the $\eta T/w$ in the confined and deconfined phases occurs. We use a Taylor expansion for the switching temperature such that: 
\begin{equation}\label{eqn:pc}
T_{sw} = T_{sw,0}\left(1-\kappa_{2}\Big(\frac{\mu_{B}}{T_{ch,0}}\Big)^{2}-\kappa_{4}\Big(\frac{\mu_{B}}{T_{ch,0}}\Big)^{4}\right).
\end{equation}

Note that this has the same format as what is used in the lattice QCD community for the chiral phase transition. The values we employ for the chiral phase transition are  \cite{Bellwied:2015rza, Borsanyi:2020fev}: $T_{sw,0}=T_{ch,0} = 158$, $\kappa_{2} = 0.0149$, and $\kappa_{4}= 0.00032$. 
However, there is no reason to assume that the minimum of $\eta T/w$ should occur exactly at the chiral transition temperature $T_{ch}$; we then study two different scenarios with a cross-over phase transition. First, we define a ``smooth" cross-over in which we identify $T_{sw}$ exactly with the chiral transition line, which in turn leads to a flat region at $T_{sw} < T < T_{min}$. Second, we contrast this with a ``sharp" cross-over where we set $T_{sw}=T_{min}$, which leads to a clear kink in $\eta T/w$ at $T_{min}$. 


\begin{table}[h!]
\begin{tabular}{ |c|c|c|c|c|c| }
 \hline
 \multicolumn{6}{|c|}{Transition Parameter Values} \\
  & Transition & $T_{sw,0}$ & $T_{ch,0}$ & $\kappa_{2}$ & $\kappa_{4}$ \\
 \hline
 CO smooth & Crossover & 158 & 158 &	0.0149 & 0.00032\\
 \hline
 CO sharp & Crossover & 199 & 158 &	0.0149 & 0.00032\\
 \hline
 \multirow{2}{5.4em}{CP(143,350)} & Crossover & 199 & 199 &	0.08 & 0.005\\
 & First Order & 158 & 158 & 0.0149 & 0.001\\
 \hline
 \multirow{2}{5.4em}{CP(89,724)} & Crossover & 199 & 199 & 0.038 & 0.0005\\
 & First Order & 158 & 158 & 0.0149 & 0.00032\\
  \hline
 \end{tabular}
 \caption{The parameters that define the pseudo-critical temperature where the minimum of $\eta T/w$ occurs, based on Eq.\ (\ref{eqn:pc}.)}
 \end{table}
 

\begin{figure*}
    \centering
    \includegraphics[width=\linewidth]{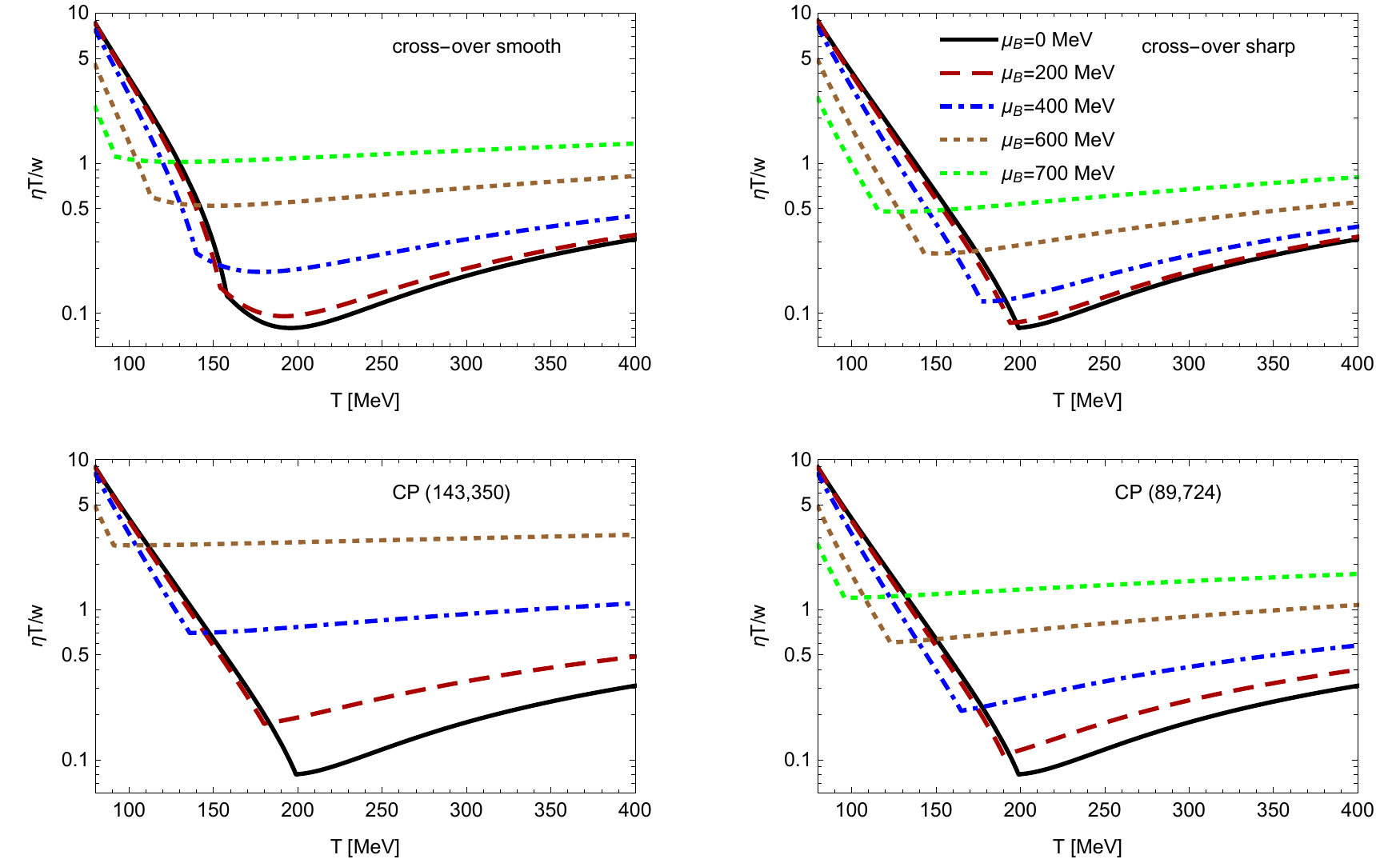}
    \caption{Temperature dependence of $\eta T/w(T,\mu_B)$ for all 4 phase diagrams along slices of constant $\mu_B$}\label{fig:etaTw}
\end{figure*}

\begin{figure}
    \centering
    \includegraphics[width=\linewidth]{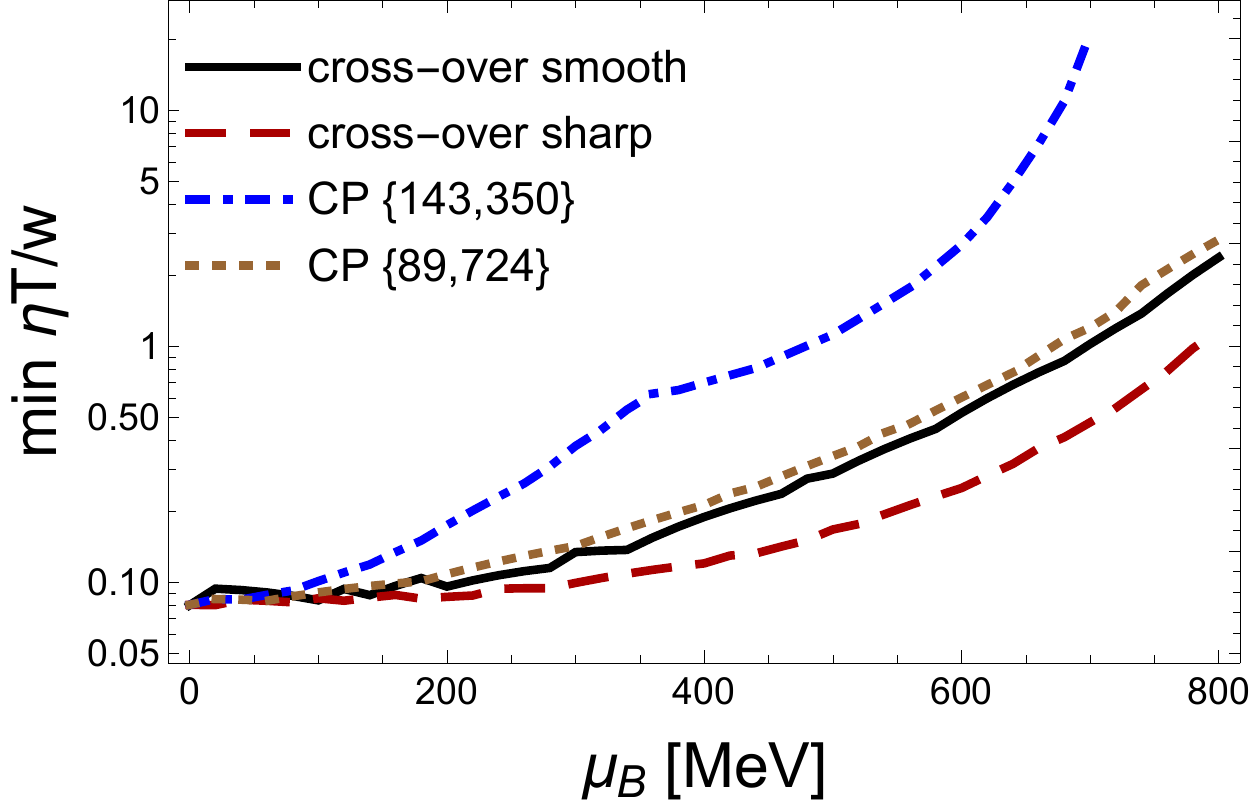}
    \caption{Minimum $\eta T/w(T,\mu_B)$ as a function of $\mu_B$}\label{fig:etaTwmin}
\end{figure}

The final results for $\eta T/w(T,\mu_B)$ are shown in Fig.\ \ref{fig:etaTw}. We can clearly see that, at vanishing chemical potential, the $\eta T/w$ has a longer flat region in the case of a ``smooth'' crossover. This small difference at $\mu_B=0$ has large effects at finite $\mu_B$ as well. In fact, the smooth crossover leads to a much larger $\eta T/w$ across all $\mu_B$ values, because the matching to the HRG occurs at lower temperatures, hence leading to a strong $\mu_B$-dependence. In contrast, matching the HRG model results at higher temperatures (as in the case of a sharp cross-over) leads to a weaker $\mu_B$-dependence, which in turn significantly decreases $\eta T/w$ at large $\mu_B$. This can be seen even more clearly in Fig.\ \ref{fig:etaTwmin}, where the minimum value of $\eta T/w$ is approximately twice as large for the smooth cross-over than for the sharp one at around $\mu_B\sim 600$ MeV. This indicates that the exact temperature for the switch from the deconfined to a confined phase has a significant effect on $\eta T/w$. In other words, the transition line for the minimum of $\eta T/w$, specifically how that varies with $\mu_B$, plays a large role in the overall magnitude of $\eta T/w$.

\subsection{Critical point}

 In the case of a critical point, we need two separate $T_{pc}$ descriptions: one for the cross-over regime and one for the first-order phase transition line. At low $\mu_B$, in the cross-over phase, we use $T_0=199$ MeV from Eq.\ (\ref{eqn:pc}), but adjust the $\kappa_2$ and $\kappa_4$ parameters to ensure that the minimum of $\eta T/w$ hits the critical point. Once the first-order phase transition is reached, we readjust the parameters in Eq.\ (\ref{eqn:pc}) such that the first-order phase transition falls onto the chiral transition line described by $T_0=158$ MeV, $\kappa_2=0.0149$, and $\kappa_{4}= 0.00032$.  
 
 
 The results for our critical point scenarios are shown in Fig.\ \ref{fig:etaTw}. The closer the critical point lies to $\mu_B=0$, the more steeply the minimum of $\eta T/w$ must drop to reach the critical point.  Since the low-T region of $\eta^{HRG} T/w$ has a stronger dependence on $\mu_B$, the magnitude of $\eta T/w$ must increase dramatically. We conclude that if a critical point is present at low $\mu_B$, the shear viscosity (at least with our setup) must be quite large. In contrast, if the critical point is far from the $\mu_B=0$ axis, such as in Ref.~\cite{Critelli:2017oub}, then the minimum of $\eta T/w$ can decrease more gradually across $\mu_B$, which in turn leads to a smaller overall $\eta T/w$ at finite $\mu_B$. It is suggestive to compare the minimum of $\eta T/w$ across $\mu_B$ for our two setups with a critical point, as shown in Fig.\ \ref{fig:etaTwmin}. We observe that the critical point at low $\mu_B$ has a significantly larger minimum $\eta T/w$.  The kink in Fig. \ref{fig:etaTwmin} arises at the critical point because the transition line changes to the chiral transition one once the first-order regime is reached.

\begin{figure*}
    \centering
    \begin{tabular}{cc}
       \includegraphics[width=0.4\linewidth]{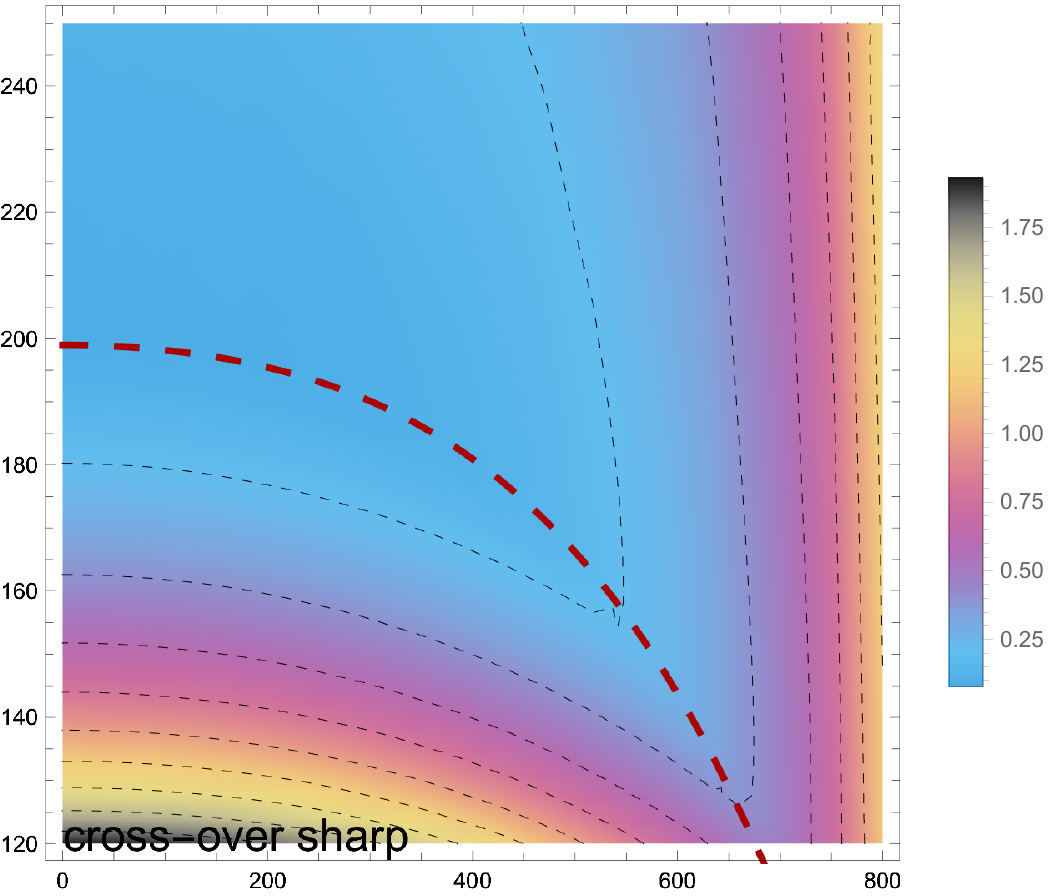}   &  \includegraphics[width=0.4\linewidth]{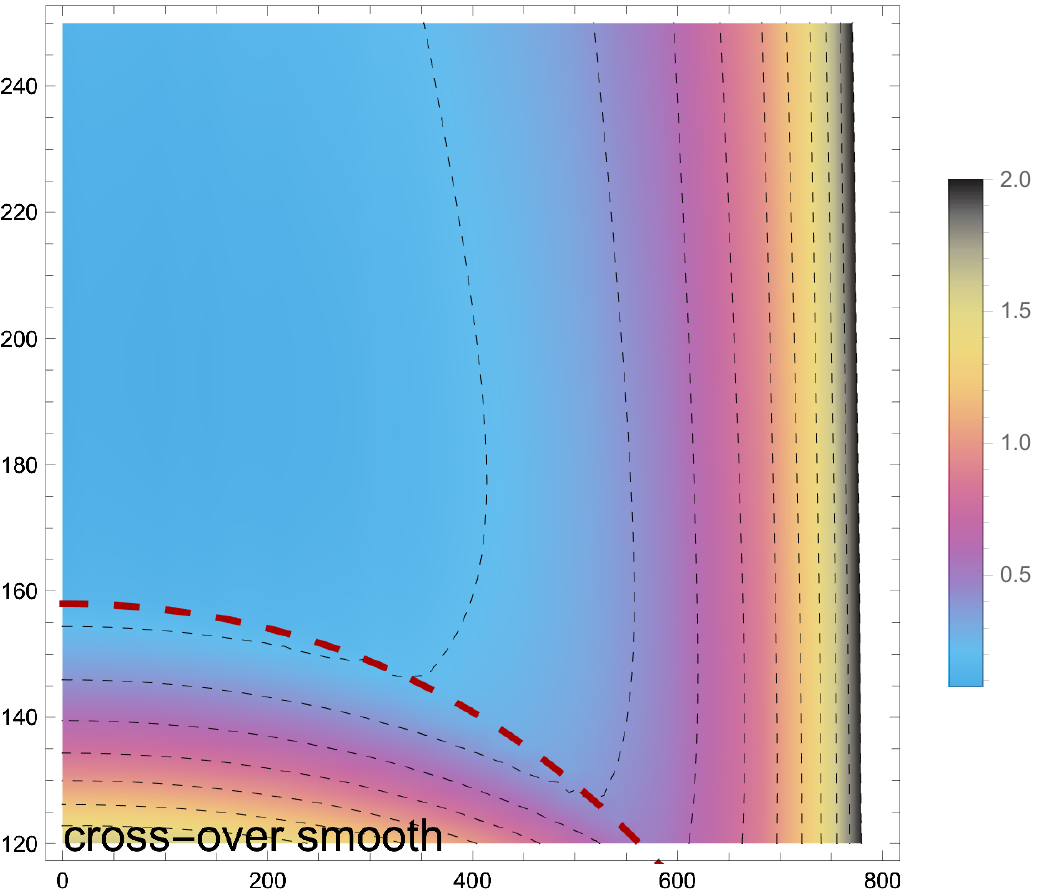} \\
      \includegraphics[width=0.4\linewidth]{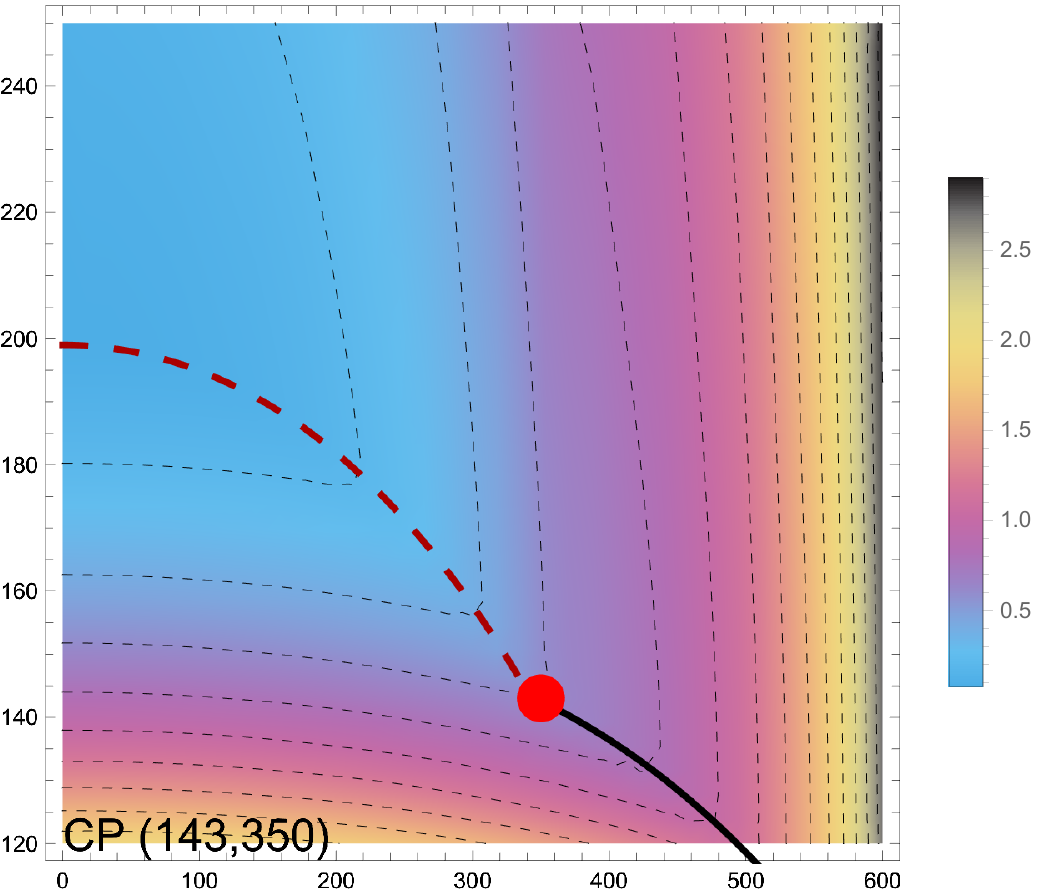}    &  \includegraphics[width=0.4\linewidth]{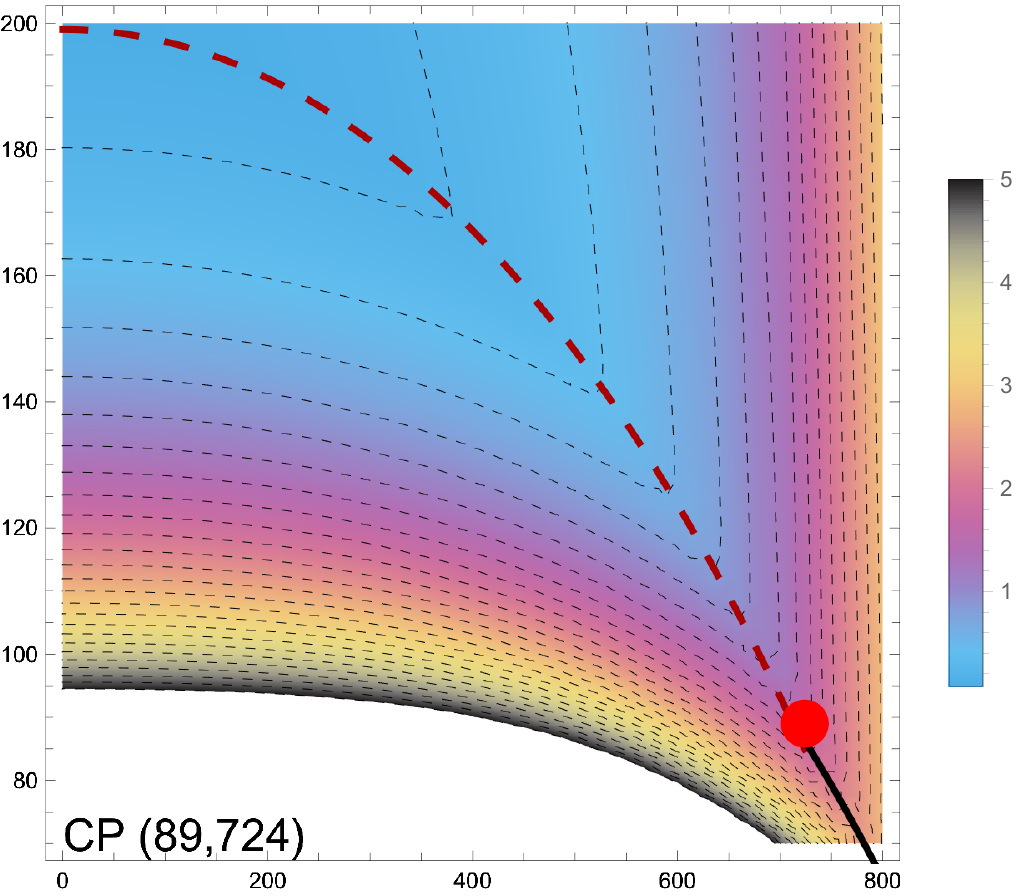}
    \end{tabular}
    \caption{Profile of $\eta T/w$ for our four different scenarios. The cross-over line is shown in red-dashed, the critical point is a red dot, and the solid black line is the first-order phase transition line. Values of $\eta T/w>5$ are shown in white.}
    \label{fig:densityplot}
\end{figure*}

Finally, in Fig.\ \ref{fig:densityplot} we show density plots of $\eta T/w$ to illustrate its different profiles. In the two cross-over setups, one can see clear differences in the trajectories of the minimum of $\eta T/w$ across the phase diagram. In the two cases with a critical point (shown as a red dot), one can clearly see the difference in the transitions lines between the cross-over and the first order phase transition line. 
 
\section{Fluctuating Hydrodynamics at \texorpdfstring{$\sqrt{s_{NN}}=19.6$}{sqrt s = 19.6} GeV}

Because of event-by-event fluctuations from the initial conditions, large variations in the local temperature in a quickly expanding and cooling QGP droplet, and entropy creation due to viscosity, it can be somewhat misleading to argue that a specific beam energy only probes a single isentropic (where the ratio of entropy over baryon number $s/\rho_B=\text{const}$) trajectory \cite{Dore:2020jye}. Rather the expanding, cooling fireball probes a range of temperatures over time as shown in Ref.~\cite{Shen:2018pty}.  Furthermore, it has been argued that applying different cuts in rapidity could provide a method for fine tuning $\mu_B$ at a fixed beam energy \cite{Brewer:2018abr}.  Thus, in order to determine the $\eta T/w(T,\mu_B)$ ranges explored at a set beam energy, we first determine the path across $\left\{T,\mu_B\right\}$ taken by AuAu collision with energy $\sqrt{s_{NN}}=19.6$ GeV.  

\begin{figure}[H]
    \centering
    \includegraphics[width=\linewidth]{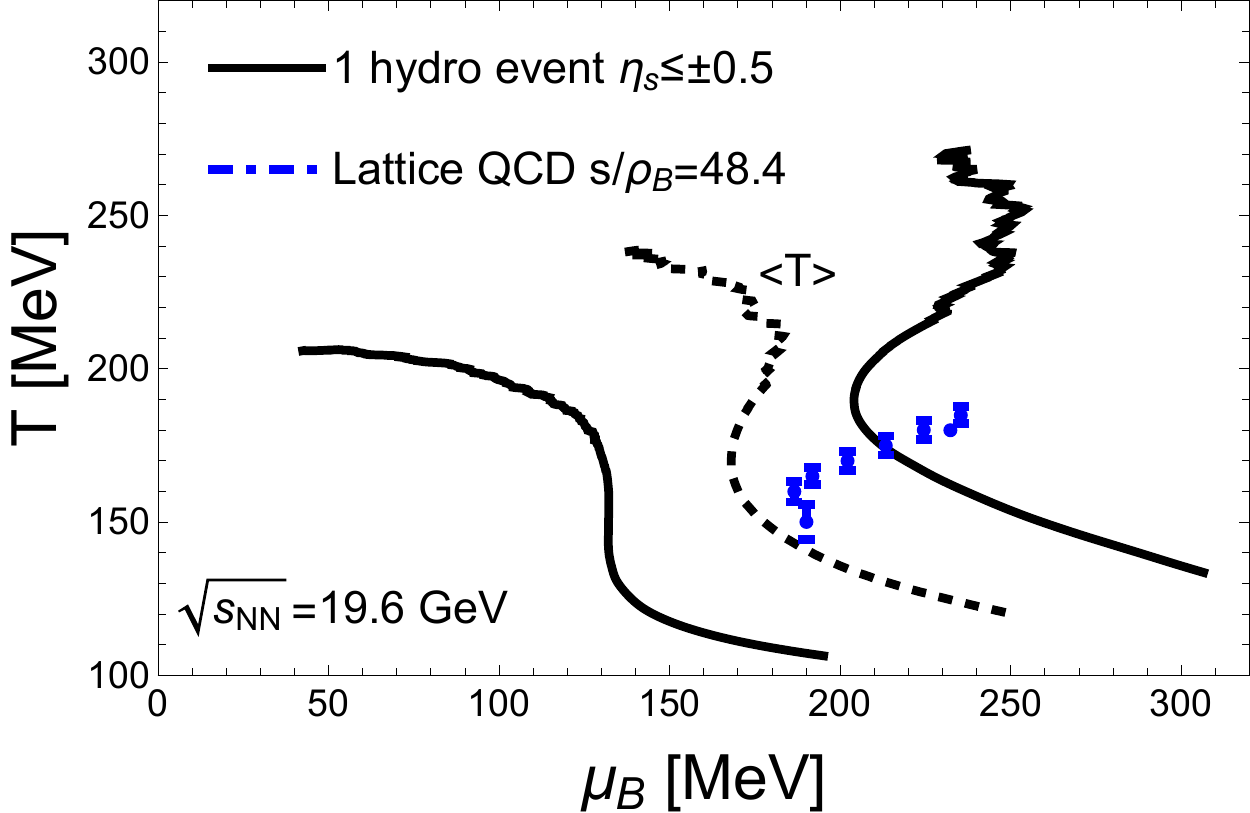}
    \caption{The temperature and chemical potential dependence of a single hydrodynamic event from Ref.~\cite{Shen:2018pty} at $\sqrt{s_{NN}}=19.6$ GeV for the mean temperature and $\mu_B$ compared to one standard deviation from the mean at mid-rapidity $|\eta_s|<0.5$. The isentrope at the same beam energy taken from lattice QCD data is also shown in blue.  } \label{fig:lat_hydro}
\end{figure}

First, we compare the commonly used isentropes calculated from lattice QCD to the $T-\mu_B$ evolution of a single hydrodynamic event from Ref.~\cite{Shen:2018pty} at $\sqrt{s_{NN}}=19.6$ GeV. From hydrodynamics it is possible to calculate the evolution of the mean temperature $\langle T\rangle$ and  chemical potential $\langle \mu_B\rangle$, which is shown in Fig.\ \ref{fig:lat_hydro}. Realistically, entropy is produced because of the finite viscosity of the Quark Gluon Plasma (both bulk and shear viscosity contribute to entropy production), so one does not expect the relationship $s/\rho_B=\text{const.}$ to actually hold.  Close to the phase transition a minimum of $\eta T/w$ is present, so it is not surprising that exactly at this point the lattice QCD isentrope most closely matches the hydrodynamic calculation. It appears that the lattice QCD isentropes can provide reasonable estimates at the phase transition, but not at higher temperatures.  

\begin{figure}[H]
    \centering
    \includegraphics[width=\linewidth]{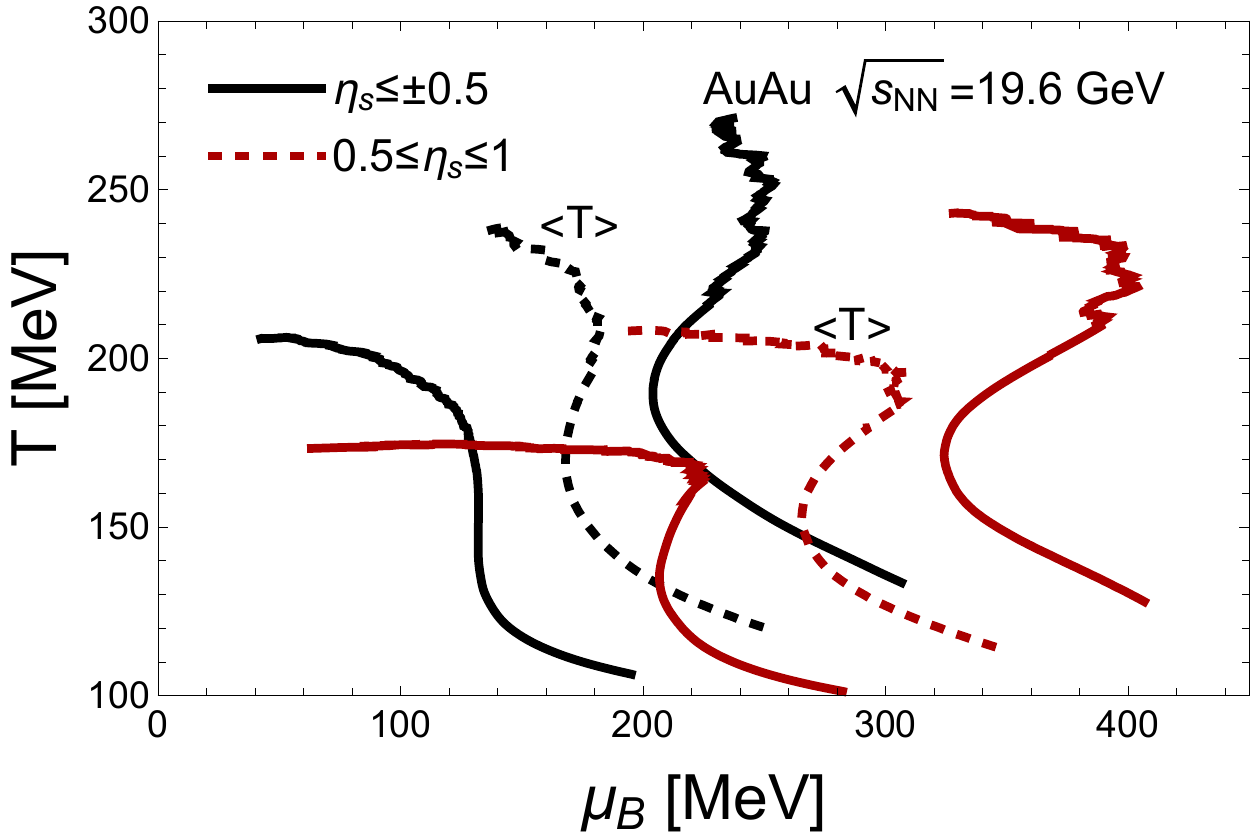}
    \caption{The temperature and chemical potential dependence of a single hydrodynamic event from Ref.~\cite{Shen:2018pty} at $\sqrt{s_{NN}}=19.6$ GeV for the mean temperature and $\mu_B$, compared to one standard deviation from the mean at mid-rapidity $|\eta_s|<0.5$ (black) and $0.5<\eta_s<1$ (red). } \label{fig:Tmub196}
\end{figure}

Additionally, in Ref.~\cite{Brewer:2018abr} it was pointed out that forward rapidity can be used to fine tune $\mu_B$ in order to refine the search for the critical point. This does indeed appear to be the case, as shown in Fig.\ \ref{fig:Tmub196}, so we expect to also see changes in the $\eta T/w(T,\mu_B)$ ranges scanned at forward rapidity. In Fig. \ref{fig:Tmub196}, two different rapidity cuts are shown: $|\eta_s|<0.5$ and $0.5<|\eta_s|<1$ (note that rapidity is typically denoted $\eta_s$, so we add the subscript to indicate spacetime rapidity and avoid confusion with viscosity). 

\begin{figure}
    \centering
    \includegraphics[width=\linewidth]{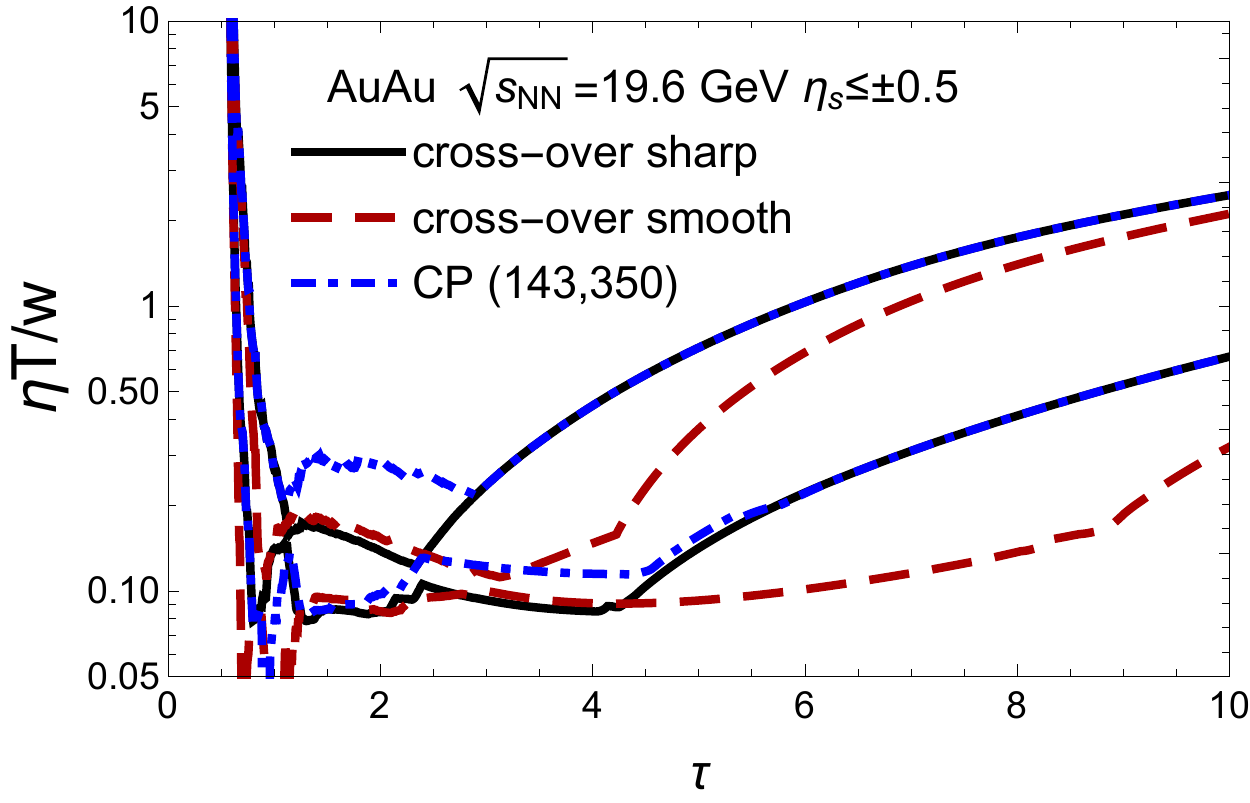}\\
    \includegraphics[width=\linewidth]{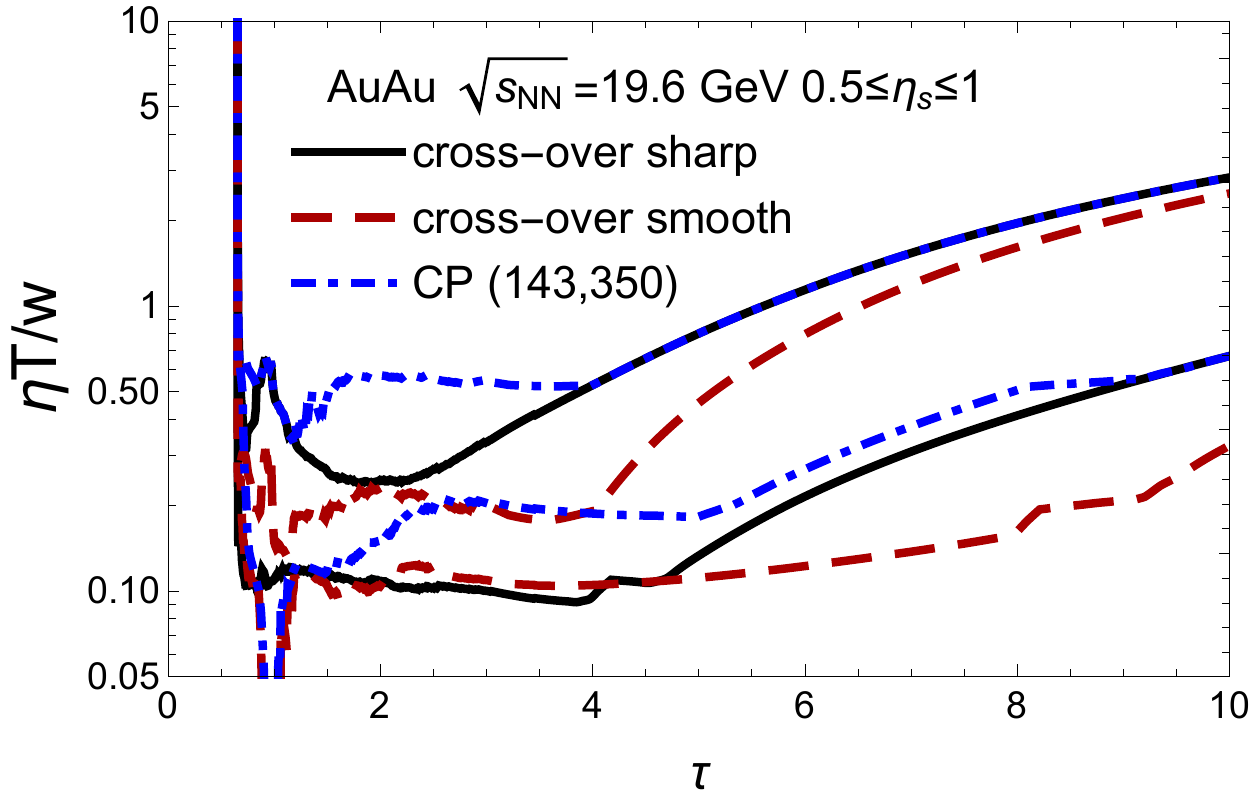}
    \caption{Approximation of $\eta T/w$  probed within relativistic viscous hydrodynamics vs. time (using the $\left(T,\mu_B\right)$ from a single hydrodynamic background from Ref.~\cite{Shen:2018pty})}\label{fig:etas_hydro}.
\end{figure}

Finally, in Fig.\ \ref{fig:etas_hydro} we compare the values of $\eta T/w$ probed across $\pm 1$ standard deviations from the mean $T$ and $\mu_B$ at the rapidity cuts $|\eta_s|<0.5$ and $0.5<|\eta_s|<1$.  Here we only compare our setups with a crossover and that with the critical point from Ref.~\cite{Parotto:2018pwx}. We find that, because the average $\mu_B$ is still relatively low (approximately $\mu_B\sim 200-300$ MeV), the location of the minimum $\eta T/w$ is relatively compatible between the different cases. Because of the longer flat region, the smooth cross-over actually sees a smaller $\eta T/w$ over a longer period of time.  In contrast, in the case with a critical point we observe a larger $\eta T/w$. We find that, at forward rapidity, the Quark Gluon Plasma sees a slightly higher $\eta T/w$ at around $\tau=1-4$ fm. Thus we expect that, while forward rapidity does reach large $\mu_B$, we also anticipate that it would be farther from equilibrium.

\section{Influence on \texorpdfstring{$T,\mu_{B}$}{T,muB} trajectories}

In this section we give a qualitative idea of the effect of using different $\eta T/w$ profiles on the trajectories taken through the QCD phase diagram by the expanding hydrodynamic system. Here, we only explore the effect of a crossover transition. The hydrodynamic setup is identical to Ref.~\cite{Dore:2020jye}, other than a replacement of the EoS with the one developed in
Ref.~\cite{Noronha-Hostler:2019ayj} with $\mu_S = \mu_Q = 0$. This is done in order to compare the effect specifically with a crossover transition. The EoS used here is directly obtained from lattice QCD through a Taylor expansion of the pressure in powers of $\mu_B/T$. The bulk viscosity has the form
\begin{equation}
         \frac{\zeta T}{w} =36\times \frac{1/3 - c_s^2}{8\pi} \, \, ,
\end{equation}
so that all of its functional dependence on thermodynamic variables enters through the speed of sound. The hydrodynamic equations exhibit Bjorken symmetry and are readily solved as a set of coupled ODE's. We compare the sharp vs. smooth cross-over form of $\eta T/w$ to either a constant $\eta T/w=0.08$ or an ideal hydrodynamic expansion. For an ideal fluid, entropy is conserved so one can describe that expansion through isentropic trajectories.

\begin{figure}
    \centering
    \includegraphics[width=\linewidth]{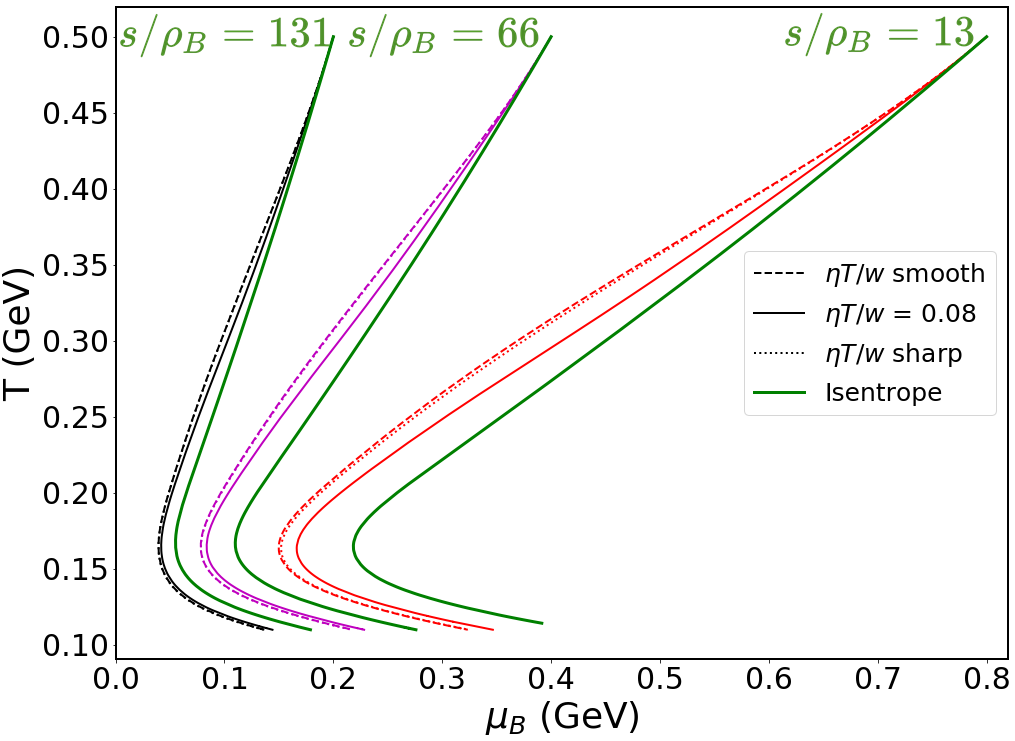}
    \caption{Comparison of $\left\{T,\mu_B\right\}$ trajectories with various initial conditions for $s/\rho_B$ with either $\eta T/w=0$ (ideal hydrodynamics), $\eta T/w=0.08$, or $\eta T/w$ following our ``sharp'' or ``smooth'' cross-over transitions.}
    \label{fig:traj}
\end{figure}

The effect on the trajectories can be seen in Fig.\ \ref{fig:traj}, where they all begin at a temperature of $T_i = 500$ MeV, and a chemical potential of $\mu_i = [200,400,800]$ MeV. The green lines are isentropes produced by running ideal hydrodynamics given the same initial conditions. The results are consistent with what is shown in Fig.\ \ref{fig:etaTw}. At lower T and $\mu_B$, the smooth and sharp shear viscosities behave very similarly, and one would expect the associated hydrodynamic trajectories to look similar if initialized in this region. Hydrodynamic runs that begin at large $\mu_B$, on the other hand, are more strongly influenced by the actual profile of $\eta T/w$, which leads to different trajectories\cite{Dore:2020jye}. 

We note that the effect of $\eta T/w$ is mainly to shift the trajectories towards $\mu_B=0$, such that a very small initial $s/\rho_B$ is necessary in order to see dramatic effects around the phase transition. For example, we reach here a similar freeze-out point with the initial condition  $s/\rho_B=13$ for both crossover $\eta T/w$ profiles, as the hydrodynamic simulation shown in Fig.\ \ref{fig:lat_hydro} where the isentropic trajectory corresponded to $s/\rho_B=48.4$. This is not surprising, because ideal hydrodynamics has maximum entropy, whereas a viscous system produces entropy over time, so that entropy production drives the system from a smaller to a larger $s/\rho_B$. 

We also compare to a constant $\eta T/w = 0.08$ to show the overall effect of having a dependence on $T$ and $\mu_B$. It can be seen in Fig.\ \ref{fig:traj} that, the higher the initial chemical potentials, the more the associated trajectories deviate from each other. It becomes apparent that, at higher $\mu_B$, one becomes more sensitive to the $T$ and $\mu_B$ dependence of $\eta T/w$ as well. Nonetheless, any viscosity (even constant $\eta T/w=0.08$) causes substantial deviations from the isentropic lines.  We conclude that heavy-ion collisions at low beam energies are more affected by the $T$ and $\mu_B$ dependence of $\eta T/w$ and we anticipate large deviations from isentropic trajectories, especially if the $\eta T/w$ profile deviates substantially from a constant value. There are some indications of this from Bayesian analyses \cite{Auvinen:2017fjw}, which imply that extreme caution should be exercised about any conclusions drawn from work based on ideal hydrodynamics at large baryon densities (see Ref.~\cite{Dexheimer:2020zzs} for further discussion). 

\section{Conclusions}

We used a hadron resonance gas picture with a state-of-the-art list of resonances to study the QCD shear viscosity at large baryon densities. We developed a phenomenological approach to produce profiles of $\eta T/w\left(T,\mu_B\right)$ on the QCD phase diagram, which can be used in relativistic viscous hydrodynamic codes simulating collisions at the RHIC Beam Energy Scan. We considered four variations: a ``smooth'' or ``sharp'' crossover, and a critical point at $\{T_c,\mu_B^c\} = \{143,350\}$ MeV or $\{T_c,\mu_B^c\} = \{89,725\}$ MeV. We have demonstrated that the trajectories followed by hydrodynamic evolution are sensitive to different behavior in the shear viscosity. It is clear that a shear viscosity with a functional dependence on thermodynamic variables may probe different areas of the phase diagram than a constant one, especially if the differences are large at early times. We note that this framework does not incorporate critical scaling in $\eta T/w\left(T,\mu_B\right)$, since this is expected to be extremely small \cite{Monnai:2016kud}, but does ensure that its inflection point (i.e. minimum of $\eta T/w\left(T,\mu_B\right)$) does converge to the critical point.

Using this framework we found that the minimum $\eta T/w\left(T,\mu_B\right)$ increases with $\mu_B$ and that this effect is strongly influenced by the exact location of the transition line one chooses for the minimum of $\eta T/w\left(T,\mu_B\right)$. For instance, a critical point at low $\mu_B$ would dramatically increase the magnitude of $\eta T/w$ compared to a critical point at larger $\mu_B$.  We showed that the trajectories from heavy-ion collisions that pass through the QCD phase diagram would have significant deviations from isentropic lines, the larger the chemical potential is. The effect of viscosity is mainly to push the trajectories towards lower $\mu_B$ at freeze-out. This implies that a significantly smaller value of $s/\rho_B$ is needed as an initial condition, in order to reach the same point of freeze-out compared to an ideal hydrodynamics system. We only consider a cross-over phase transition for this particular part of the study but the presence of a critical point would only enhance this effect due to the critical scaling of bulk viscosity, as shown already in Ref.~\cite{Dore:2020jye}. We also expect early times ($\tau\sim 1-4$ fm/c) to be the regime probing the minimum of $\eta T/w$.










\section{Acknowledgements }

This material is based upon work supported by the National Science Foundation under grant
no. PHY-1654219 and by the U.S. Department of Energy, Office of Science,
Office of Nuclear Physics, within the framework of the Beam Energy Scan Theory (BEST) Topical
Collaboration. We also acknowledge the support from the Center of Advanced Computing and
Data Systems at the University of Houston.  J.N.H. acknowledges support from the US-DOE Nuclear Science Grant No. DE-SC0020633 and from the Illinois Campus Cluster, a computing resource that is operated by the Illinois Campus Cluster Program (ICCP) in conjunction with the National Center for Supercomputing Applications (NCSA), and which is supported by funds from the University of Illinois at Urbana-Champaign.  E.M. was supported by the National Science Foundation via grant PHY-1560077. P.P.  acknowledges support by the DFG grant SFB/TR55.

\nocite{*}
\bibliography{all}

\end{document}